\documentclass[10pt]{article}

\usepackage[dvips]{epsfig}
\usepackage{amsmath}
\usepackage{amssymb}
\usepackage{cite}
\usepackage{lineno}
\usepackage{microtype}
\DisableLigatures[f]{encoding = *, family = * }
\usepackage{rotating}

\usepackage[labelfont=bf,labelsep=period,justification=raggedright]{caption}
\makeatletter
\renewcommand{\@biblabel}[1]{\quad#1.}
\makeatother

\begin{document}


\begin{center}
{\Large
Astronomical Alignments of
the Sun Temple site in Mesa Verde National Park
}
\end{center}
\vspace*{0.5cm}

\begin{center}
Sherry Towers$^{1,\ast}$
\\[0.5cm]
$^{1}$ Simon A. Levin Mathematical, Computational and Modeling Sciences
Center,\\ Arizona State University, Tempe, AZ, U.\,S.\,A.
\\[0.25cm]
$^\ast$ Correspondence to: {\tt smtowers@asu.edu}
\end{center}


\vspace*{0.5cm}

\section*{Abstract:}

Summer 2015 marks the 100$^{\rm th}$
anniversary of the excavation by J.\,W.~Fewkes of 
the Sun Temple in Mesa Verde National Park, Colorado; an ancient ceremonial complex of unknown
purpose, prominently located
atop a mesa, constructed
by the Pueblo Indians approximately 1000 years ago.  
In this analysis we perform a digital survey of the site, and
examine the possibility that four key tower-like elements of the complex
were used for observation of the rise or set of celestial bodies known to
be sacred to the Pueblo Indians.

We find statistically significant evidence that the site was used for astronomical observation
of the rise and/or set of nearly all such bodies.
The Sun Temple appears to represent the most comprehensive prehistoric astronomical 
observatory yet uncovered.

\section*{Main Text:}

\section{Introduction}

Ample evidence exists throughout the world that many prehistoric societies made detailed observations of
the heavens;
structures such as Stonehenge~\cite{hawkins1963stonehenge,hawkins1968astro,thom1975stonehenge},
and other prehistoric sites in Europe, the Americas, Africa, India, and Asia,
have been shown to have key alignments
to the yearly solstitial cycle of the Sun,
and also the $18.6$ year cycle the Moon when
it reaches its most northerly and southerly rise and set on the horizon, known
as the lunar standstills \cite{hawkins1963stonehenge,hawkins1968astro,thom1971megalithic,thom1975stonehenge,aveni2003archaeoastronomy,kelley2005exploring}.
The potential of alignments to the horizon rise and set points
of bright stars at many of these sites has not yet been comprehensively studied,
but it likely would also have been obvious to the ancients (as it is to us) 
that some stars are much brighter than others,
their relative positions in the sky appear to be fixed during our lifetimes, most have
seasonality of their visibility in the night-time sky, and 
they always appear to rise and set at the same point on the horizon.
The solstice cycle of the Sun and the seasonality 
of the night-time visibility of these bright stars 
would likely have helped to anticipate the time of
planting and harvesting and other notable periods of the year, forming the basis of the first calendars.

Most of the ancient observatories so far uncovered predate written history, and
oral lore regarding the sites has been lost.  In these cases, we thus must rely upon the physical
evidence left by the ruins themselves to try to determine as much as we can
about their intended purpose.
In the study presented here, we examine the Sun Temple at Mesa Verde National Park in Colorado,
an ancient ceremonial complex built around 1000 years ago by the Pueblo Indians.

Mesa Verde covers an area of over 210 km$^2$, and the park topography consists of a
series of many small mesas, separated by deep side canyons \cite{wenger1987story}.
The area was
settled by Pueblo Indians beginning around
around 470 AD,
with final abandonment in late 1200's due to
drought conditions \cite{stiger1979mesa,adler1996ancestral,cordell2007mesa,varien1999sedentism,van2000environmental}.
Several thousand ruins associated with this period of
occupation are found throughout the park~\cite{glowacki2015living},
with the
most famous structure being the Cliff Palace, which is built into a cliff 
underneath a large rock overhang.
The Cliff Palace site was 
first inhabited in the mid-1000's AD, and 
finally abandoned in the late 1200's~\cite{getty1935new,gibbon1998archaeology}.

\begin{figure}[h]
   \begin{center}
    \mbox{\put(-190,0){ \epsfxsize=13cm
           \epsffile{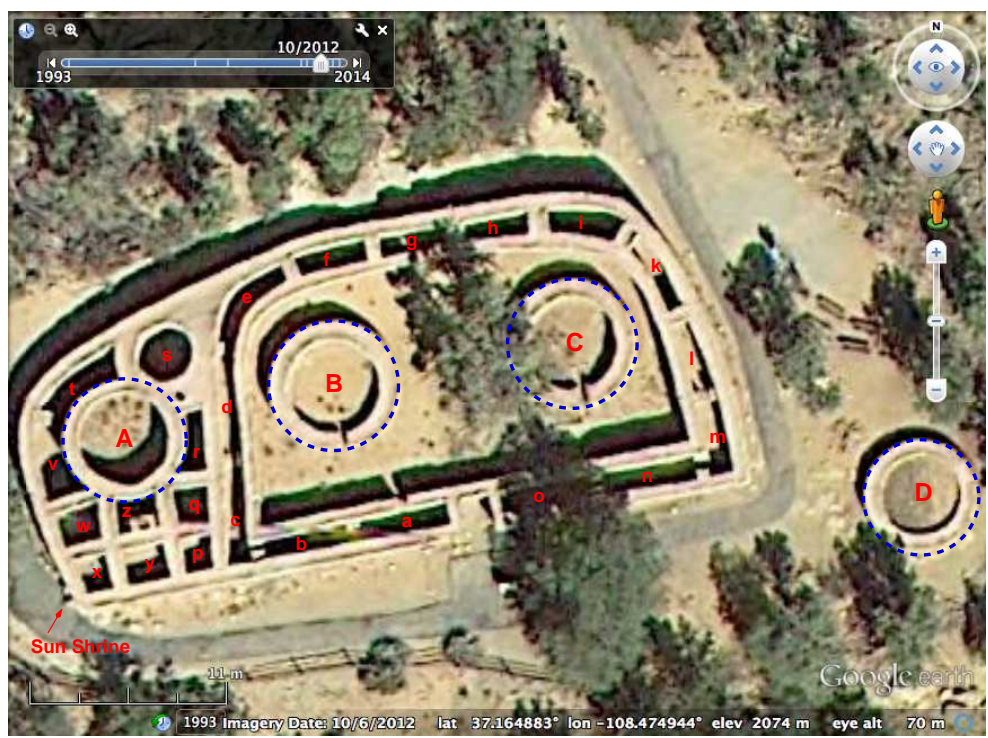}
     }}
     \vspace*{-0.0cm}
  \caption{
      \label{fig:sun}
Aerial view of the Sun Temple complex 
(as obtained from Google Earth,
accessed February 1, 2015), 
with sections labeled according to
\cite{fewkes1916excavation}, and the location of the Sun Shrine indicated (a
naturally eroded star-shaped basin).  
The ground length over the width of the view is
just over 56 meters.  Overlaid are circles over the four kiva-like structures.
The different lines that can be drawn tangent to two kivas, or tangent
to one and transecting another, or transecting both, are considered as potential astronomical sight
lines in this analysis.
   }
   \end{center}
 \end{figure}

Directly across the canyon from Cliff Palace is the Sun Temple,
built atop a mesa with a commanding view of the surrounding landscape.
An aerial view of the Sun Temple ruin is shown in Figure~\ref{fig:sun}.
The D-shape of the complex is recognized
by modern Pueblo Indians to denote a ceremonial structure, however
information regarding the exact use of such structures has been lost in oral traditions.  
Indeed, the complete lack of domestic artifacts and trash mounds associated with the site point to
its use for ceremony, rather than habitation, and the
site is extraordinarily unique in the region in this respect, and
also in its architecture~\cite{fewkes1916excavation,munson2011legacy}.  

Previous studies have shown that key architectural features in the Cliff Palace
have solar solstice and lunar 
standstill alignments with the Sun Temple~\cite{malville1993prehistoric,munson2010reading},
and a nearby complex, Balcony House, has been shown to have been used for
observations of the Sun and Moon~\cite{fiero1998balcony}.
However, to-date there has been no
comprehensive study of the potential use of the Sun Temple by itself as an observatory.

A notable feature of the Sun Temple complex is the incorporation of
four large walled circular structures, which, following Fewkes, we shall refer
to as ``kivas'' (even though, apart from their circular walls, 
they deviate in many respects from the usual form of a kiva in traditional
Pueblo architecture~\cite{munson2010reading,mesaverde2011notes}, and in some respects
their original form was more akin to a typical tower~\cite{munson2010reading}).
These structures are indicated on the aerial view of the Sun Temple in Figure~\ref{fig:sun}.
Based on the masonry patterns, 
these structures have been posited to have preceded the construction of the remainder of the
complex, but by an unknown period of time~\cite{munson2010reading}.
In this analysis, we assume that because they were constructed
first, these kivas played a key role in the function of the temple,
and we thus consider site alignments going either tangent to, and/or through, pairs
of these four kivas as potential sight lines used for astronomical observations of celestial
bodies rising and setting on the horizon.
We posit that lines tangent to pairs of kivas could have been used as ``gun sights'' to precisely locate
positions on the horizon,
In addition, slits in a kiva can be used to locate positions on the
horizon, particularly when aided by a tangent line to one or more additional kivas.  
Wall slits used as astronomical
viewing aids have been documented in other 
Pueblo architecture~\cite{malville1993prehistoric,macey2013encyclopedia,penprase2011celestial,munson2010reading,reyman1975nature},
and the use of paired 
features as ``gun sights'' for astronomical observation by the Pueblo has also been documented~\cite{malville1993prehistoric}.

Pueblo Indians consider the Sun, Moon and Venus to play a sacred role in their cosmology, along with the
Pleiades, the seven bright stars that we identify with the constellation 
Orion~\cite{parsons1926ceremonial,bunzel1932introduction,ellis1975thousand,malotki1983hopi}, and
several other bright stars~\cite{miller1997stars}.
In this analysis we focus only on potential alignments to these sacred celestial bodies.
As a cross-check to the analysis, we also examine the potential of alignments to celestial bodies
not considered sacred to the Pueblo.  

Using topographical information on a 30 m grid from the Shuttle Radar Topography Mission (SRTM), 
we determine the horizon profile at the site, and using the PyEphem library of ephemeris calculation 
software\footnote{Available from {\tt http://pypi.python.org/pypi/pyephem}, accessed
January 30, 2015.}
we determine the 
declinations
of all stars that are known to be sacred to the Pueblo people, around the
time the Sun Temple is presumed to have been built (circa 1250 AD~\cite{glowacki2015living}).
We also determine the declinations of
the Sun at equinox and summer and winter solstice, and the Moon at major and minor northern and
southern standstills, and the maximum
and minimum declinations of Venus as a morning and evening star.

Obviously, with several site lines, and several potential astronomical alignments, the
probability of finding matches just by mere random chance cannot be ignored. 
We reduce, as much as possible, the probability of spurious matches by only considering
alignments between the four key features of the site, and only considering celestial bodies
that are considered sacred to the Pueblo Indians.
This problem of spurious matches has unfortunately often been overlooked by many past researchers, and several 
claims of
purported prehistoric astronomical
observatories have later been shown to have no supporting statistically significant evidence
of alignments (see, for instance, the discussion
in References~\cite{hawkins1974astronomical,hawkins1975astroarchaeology,haack1987critical}).
We present here a method for testing 
the null hypothesis that the Sun Temple site was {\it not} used as an observatory by using
stochastic methods to randomly rotate the configuration of the
Sun Temple complex over many repetitions, and thus determine
the probability distributions for the expected number of site alignments randomly matched to stars, and the
expected number of stars randomly matched to site alignments.  

As we will additionally discuss, as a cross-check of the analysis
we also examine potential alignments to celestial bodies up to visual
magnitude 4 that are {\it not} considered sacred
to the Pueblo Indians.

In the following sections we describe the Sun Temple complex and the methodology used to 
determine the site alignments, the rise/set azimuths of celestial bodies, and the criteria used
to declare a potential astronomical alignment match, along with a description of the methods
used to assess the statistical significance of the number of observed matches.

\section{Methods and Materials}
\subsection{Description of the Sun Temple}

The Sun Temple is situated west of Cliff Palace, on the promontory of Chapin Mesa
 formed by the confluence of Cliff and Fewkes Canyons. 
As described by the National Park Service~\cite{mesaverde2011notes}:
\begin{quotation}
The ruin was purposely constructed on a commanding promontory in the neighborhood of large inhabited cliff houses. It sets somewhat back from the edge of the canyon, but near enough to make it clearly visible from all sides, especially the neighboring mesas. It must have presented an imposing appearance rising on top of a point high above inaccessible, perpendicular cliffs. No better place could have been chosen for a religious building in which the inhabitants of many cliff dwellings could gather and together perform their great ceremonial dramas.
\end{quotation}

When Jesse Walter Fewkes excavated the Sun Temple
site in the early 1900's, he noted that the walls
were made of fine, carefully pecked masonry blocks,
were exceptionally vertical, and that their original
height was likely around two meters above the present height~\cite{fewkes1916excavation}. 
He also noted that the complex had had no roof, and erroneously 
concluded that the structure had never been completed~\cite{fewkes1916excavation,munson2010reading}.
Fewkes repaired parts of the complex (for instance, by installing capping
on the walls to prevent erosion), but from a recent survey of the site and the photographs
of the excavation process, it has been concluded that Fewkes did not change the layout
of the site~\cite{munson2010reading}.

An aerial view of Sun Temple is shown in Figure~\ref{fig:sun}.  
The width of ruin at widest is around 64 feet, and length is 122 feet.
The walls are on average 4 feet thick, made of masonry surrounding rubble core.  

The site has proven difficult to date, largely because of the lack of artifacts and wood
for dendrochronological dating; based almost entirely upon its geographic proximity to the Cliff
Palace, the site has been presumed to have been constructed in the 
1200's~\cite{fewkes1916excavation,
munson2011legacy,glowacki2015living}. 
However, the use of the pecked block core-and-veneer ``McElmo'' style masonry seen
in the Sun Temple
complex has been dated elsewhere in the region to have been used between the early 1100's to the 
1200's~\cite{van2010connecting}, thus an earlier construction date is possible~\cite{munson2010reading}.

There is a notable ground feature on the southwest corner between two short walls that
jut out from the side of the complex
to either side of a naturally eroded star-shaped basin approximately two feet across.
Fewkes dubbed
this feature the ``Sun Shrine'' (see Figure~\ref{fig:sun}).
Because of this feature, 
he believed that the complex was used for Sun worship, and
claimed that there was a solstice alignment along the south wall of the complex, and that the
Sun Shrine on the southwest corner could have been used to observe the winter solstice sunset.  
However, it was pointed out in Reference~\cite{reyman1977solstice} that the alignment of the south wall is several degrees different from any solstitial or equinoctial alignment, and no apparent use of the Sun Shrine for
solstice observation has been found~\cite{malville1993prehistoric}.
As we will discuss, while we do not explicitly include the Sun Shrine in our analysis, 
we do find that two key astronomical 
alignments at the site go through it, indicating its apparent special significance (although
not for solar solstice observation).

Also notable in the Sun Temple complex are four circular
structures, superficially similar to ceremonial Puebloan structures called kivas. 
Unlike typical kivas, however, these are constructed with their bases at ground level rather than below ground, 
and lack a fire pit, air deflector, and
bench~\cite{munson2010reading}. Three
of these kiva-like structures are within the main structure, and the 
fourth is a tower that lies outside.
These four features are marked in the aerial view of the 
site shown in Figure~\ref{fig:sun}.  Following Fewkes, we refer to them
as Kivas A, B, C, and D.

The use of the Sun Temple for astronomical observation of the Sun winter solstice set and major Moon
southern standstill set, {\it as viewed from across the canyon from Cliff Palace}, has been previously
noted by Malville~\cite{malville1993prehistoric}.  However, potential astronomical alignments as viewed from within
the Sun Temple itself have had yet to be considered.

\subsection{Site Survey}

Google Earth is a virtual globe, map, and geographical information systems (GIS)
program, freely available from {\tt http://earth.google.com} (accessed
January 30, 2015).
Since the launch of the product in 2005, it has been used in a wide range
of academic endeavors, including for use in the survey of
archaeological sites 
(see, for instance, References \cite{ur2006google,zhao2007introduction,myers2010field,sadr2012google,kennedy2011google}).
In this analysis we use Google Earth to obtain satellite imagery and geographic information related
to the Sun Temple site.
We also used Google Earth for cross-checking our calculations of the
orientation of site alignments, the horizon profile, and the relationship of the Sun Temple to the surrounding
landscape.

To survey the site, an aerial view of the site was obtained from Google Earth, including the image distance scale
(see, for instance, Figure~\ref{fig:sun}).  The image was then read into Xfig, a free and open-source vector graphics 
program\footnote{See {\tt www.xfig.org}, accessed March 1, 2014}. 

Within Xfig, circles were overlaid
onto the four kiva-like structures, and the radii and centers of the circles in the coordinate frame of
the image were determined.  Because there is some amount of objectivity involved in the placement of the circles, the procedure
was repeated ten times, and the average and one standard deviation uncertainty on the kiva centers and radii determined
from the ten iterations.
The precision of this benchmark placement procedure was cross-checked by applying the process to an
aerial view of the 100 m Olympic track 
at the Crystal Palace National Sports Centre in London; an average precision of approximately $\pm15$ cm was obtained,
in concordance with the estimated precision of most of the benchmarks overlaid at the Sun Temple site.
The precision of the benchmark placement procedure tends to be better than the pixel-size of the aerial image
itself when lineal features are present with marked color gradations along the line.  Both the Sun Temple image and
the image of the Olympic track have such lineal color gradations.

Taking pairs of kivas, we determined all lines that go either tangent to both kivas, through both kivas, or tangent to
one and through the other.
The only tangents to pairs of kivas that were considered for this
analysis were ones that formed a ``gun sight'' line of sight between the two kivas.

\begin{figure}[h]
   \begin{center}
    \mbox{\put(-190,0){ \epsfxsize=13cm
           \epsffile{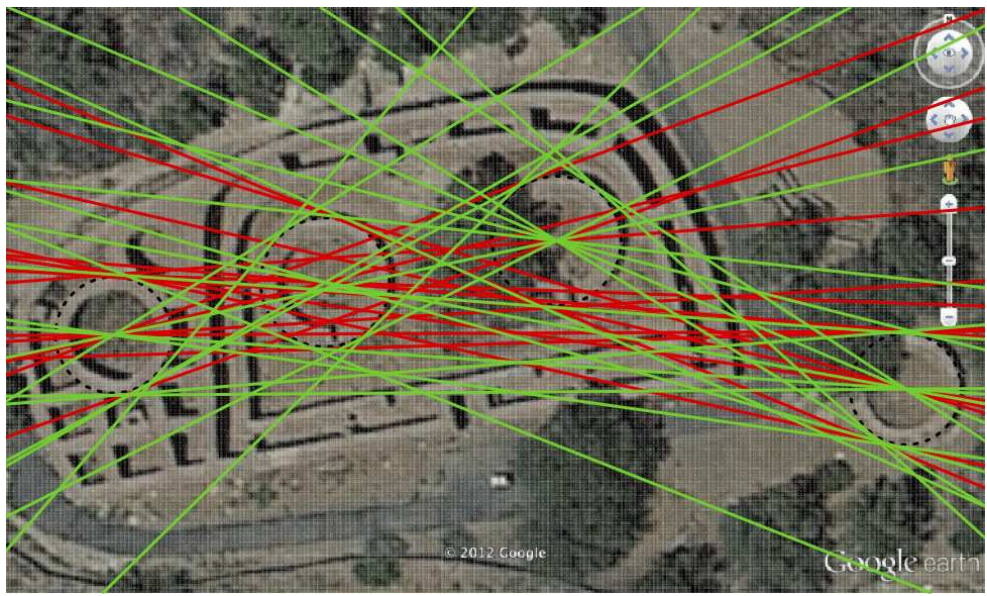}
     }}
     \vspace*{-0.0cm}
  \caption{
      \label{fig:all_lines}
All site lines that go either tangent to pairs of the four kivas, tangent
to one and through the center of another, or through the center of both.
Shown in red are the site lines that cross one of the other kivas more than
a foot from the center, which would
make sighting along that line impractical because portholes
would be difficult to place in the masonry at such oblique
angles. 
Only the green lines are considered in this analysis.
Aerial view obtained from Google Earth
(accessed February 1, 2015). 
   }
   \end{center}
 \end{figure}

All such lines are shown in Figure~\ref{fig:all_lines}.  Note that some of the lines
cross other kivas in ways that, if all the kivas were of the same height, would interfere
with the line of sight because portholes would be
impractical to place in the kiva walls at oblique angles.  
We thus take a conservative approach, and exclude lines that cross through another
kiva further than a foot from its center.  This leaves a total of 22 site alignments to
consider.

\subsection{Horizon Profile}

Determination of the rise and set azimuths of celestial bodies on the horizon requires knowledge of the horizon profile. 
 The Shuttle Radar Topography Mission (SRTM) recorded elevation data for most places on Earth,
 on a grid with tiles spaced
 30 meters apart. The information is available from {\tt http://srtm.csi.cgiar.com} (accessed January 30, 2015).
Using the methodology of~\cite{yoeli1985making}, we used this information to
determine the view shed of the site and the horizon profile.
We assumed that the height of the walls of the complex were two meters above their current height~\cite{fewkes1916excavation}.

Based on the horizon profile, in small steps of azimuth of $0.01^\circ$, we created a
look-up table of the declination of a hypothetical celestial
body rising at that particular azimuth. We also calculated the declination of a celestial body
setting at the azimuth pointing in the opposite direction.

\subsection{Celestial body rise/set azimuths as viewed from Sun Temple}

Because of gradual changes in the Earth's axis of rotation (called precession), in addition to
motion of the stars relative to our own solar system (know as proper motion),
the rise/set azimuths of stars change slowly over time~\cite{hawkins1968astro}.
The coordinates of a star in the Earth's equatorial coordinate system are known as the declination, $\delta$, and right
ascension,
$\alpha$, and
the cumulative change in these coordinates over time can be significant for some stars, changing their
rise/set azimuths by several degrees over a few centuries.

Using the PyEphem library of ephemeris calculation software,
we calculated the 
declination and right ascension of the catalog of all stars listed in the PyEphem catalog up to visual magnitude of 4.
We calculated these quantities at dates separated by 25 years from 1000 AD to 1400 AD. Based
on masonry style of the Sun Temple, 
dates before 1100 AD are implausible~\cite{van2010connecting,munson2010reading,varien1999masonry}, 
and the region was abandoned due to severe drought by the late 1200's AD~\cite{glowacki2015living,varien1999masonry,varien1999sedentism,van2000environmental}).
We examine the earlier and later dates as a cross-check; if indeed the complex was used to observe
sacred celestial bodies, the largest number of site line matches to the rise and set of those bodies (and
vice versa) should occur near the date of the site construction.

At each date range considered, we also calculated the declination of the Sun at its solstices, and the
Moon at major and minor standstills.
The declination of Venus as a morning and evening star can change significantly over the course of just
a century. For each 25 year time span we used the temporal maximum and minimum 
declination information for Venus provided in Reference~\cite{vsprajc2015alignments},
as both a morning and evening star. 

Based on the declination, and horizon profile, we calculated the rise/set azimuth of each star at
each date considered.  

\subsubsection{Celestial bodies considered sacred by the Pueblo Indians}

Pueblo Indians consider the Sun, Moon and Venus to play a sacred role in their cosmology, along with the
Pleiades, and the seven bright stars of Orion (Rigel, Betelgeuse, Bellatrix, Alnilam, Alnitak, Saiph, and 
Mintaka)~\cite{parsons1926ceremonial,bunzel1932introduction,ellis1975thousand,malotki1983hopi}. They
have a complex ceremonial calendar believed to have been originally driven by observations of
celestial bodies (but later largely supplanted by the Gregorian calendar, under pressure from
later Spanish influence)~\cite{tedlock1983zuni,malotki1983hopi,mccluskey1977astronomy}.
Major ceremonies are centered around the Sun winter solstice in particular, and observations of
the Sun along the horizon 
predominantly (but not exclusively) take place at sunrise~\cite{zeilik1985ethnoastronomy,zeilik1991historic}.
Reference~\cite{miller1997stars} also identifies the four stars in the great square of Pegasus (Sirrah,
Markab, Algenib, and Scheat), along with Deneb, Vega, Arcturus, Spica, Antares, Sadr, and Albereo,
as being sacred to the Pueblo Indians.
We consider the Moon at each of its major and minor northern and southern standstills, 
the Sun at the summer and winter solstice and equinoxes, and Venus at its most northern and southern
declinations as a morning and evening star to be separate sacred bodies (nine in total).

In this analysis we focused on potential site alignments to these celestial bodies.
As a cross-check, we examined the number of site alignments to celestial bodies with visual
magnitude less than 4 that are not in this list (ie; celestial bodies not known to be sacred to the Pueblo
Indians).  We also performed this cross-check for such bodies with visual magnitude less than 3, and less than 2.

\subsection{Assessment of Statistical Significance of Number
of Observed Alignments}

Given $M$ site alignments, and $N$ celestial 
bodies,
we considered a site alignment
to potentially have been used for observation of a celestial body if the rise or set declination of the site alignment
is within a degree of the 
declination of the body.
We examined all $M\times N$ possible potential matches of celestial bodies and site alignments and determined the fraction
of site alignments that were matched to within a degree of the declination of the celestial body, and vice versa.

In order to test the probability of obtaining the observed number of such matches, we performed 10,000 iterations
wherein we randomly rotated the Sun Temple complex by an angle uniformly randomly sampled between 0 and 180 degrees.
For each iteration we determined the
number of matches of site alignments to stars, and vice versa.  From these simulations, we determined
the probability of obtaining at least as many matches as we actually observe, under the null hypothesis
that the complex was not used for astronomical observations of the celestial bodies considered.

\section{Results and Discussion}

\begin{figure}[h]
   \begin{center}
    \mbox{\put(-190,0){ \epsfxsize=13cm
           \epsffile{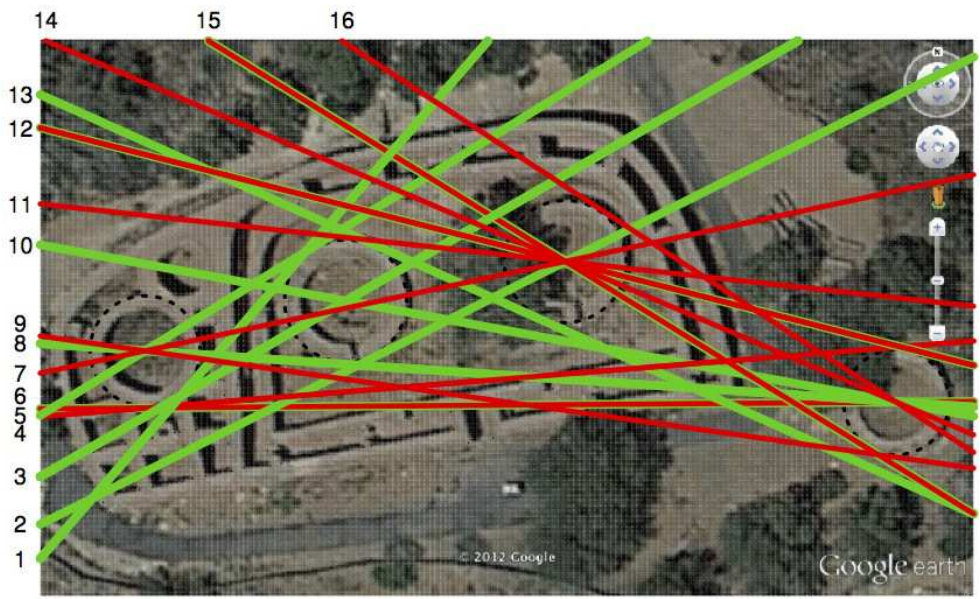}
     }}
     \vspace*{-0.0cm}
  \caption{
      \label{fig:matches}
Site lines matched to the rise or set azimuth of celestial bodies considered sacred to the Pueblo
Indians.  Green (red) lines indicate lines apparently used to observe the rise (set) of the celestial bodies
in 1250 AD (note that some lines are used for both).
The westernmost kivas are primarily associated with the observation of the rise of sacred celestial bodies,
while the easternmost kivas are primarily associated with the observation of the set.
The names and declinations of the celestial bodies 
corresponding to the line numbers are summarized in Table~\ref{tab:tab1}.
Aerial view obtained from Google Earth
(accessed February 1, 2015). 
   }
   \end{center}
 \end{figure}

In Figure~\ref{fig:matches} we show the site alignments that matched to the rise
and/or set of the sacred celestial bodes. 
The declinations and rise and set azimuths of these bodies
in 1250 AD are shown in Table~\ref{tab:tab1}.
Interestingly, and perhaps significantly, all but two of the nine site alignments matched to the
set of sacred celestial bodies are associated with Kiva D, and five
of the nine go through the center of Kiva C.
Of the ten lines associated with the rise of sacred celestial bodies, all but three 
are associated with Kiva B, and those three are the only ones that also serve as alignments
to the set of other celestial bodies.  The westernmost kivas thus seem to be primarily
associated with the observation of the rise of sacred celestial bodies, while the easternmost kivas
are primarily associated with the observation of the set.

A total of 86\% of the 28 sacred bodies examined had a 
declination matched to within a degree of a site alignment at 1250 AD, and
86\% of the 22 site alignments had a match to a sacred celestial body (combined $p=0.0023$).  

As a cross-check, we examined the celestial bodies in the PyEphem star catalog up to magnitude 4
that are not known to be sacred
to the Pueblo people; only 58\% had a rise/set match to a site alignment, which
is not statistically significant ($p=0.27$). 
Examination of non-sacred
stars in the catalog only up to visual magnitude 3, and also only up to visual
magnitude 2, also did not yield a significant number of matches to site alignments
($p=0.41$ and $p=0.56$, respectively).

We thus conclude that the site does indeed appear to have been used for observation of
celestial bodies known to be considered sacred, 
but there is no statistically significant evidence
that it was used for observation of other stars.

\begin{figure}[h]
   \begin{center}
    \mbox{\put(-190,0){ \epsfxsize=13cm
           \epsffile{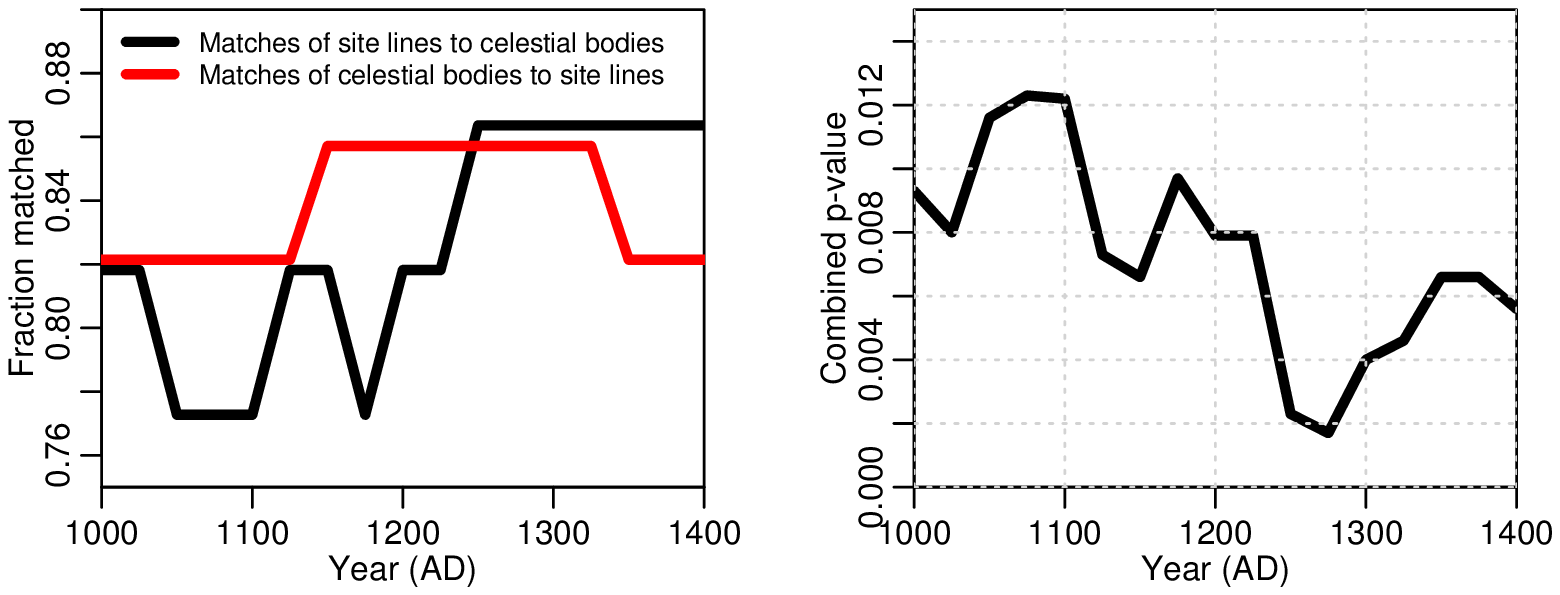}
     }}
     \vspace*{-0.0cm}
  \caption{
      \label{fig:date}
The left hand plot shows
the fraction of sacred celestial bodies with rise or set matched to a site line vs hypothesized date of construction
of the Sun Temple (red).  
Also shown is the fraction of site lines that are matched to the rise or set
of a sacred celestial body vs date (black).
The right hand plot shows the combined probability, under the null hypothesis that the complex was
not used for astronomical observation of sacred bodies, of observing at least as many site line to star
matches, and vice versa, as actually observed (the smaller the p-value, the more likely the date
of construction).
Based on the number of alignments, construction dates before 1250 AD appear to be more unlikely,
but cannot be statistically excluded.
   }
   \end{center}
 \end{figure}

In Figure~\ref{fig:date} we show
the fraction of sacred celestial bodies with rise or set matched to a site line vs hypothesized date of construction
of the Sun Temple.  
Also shown in Figure~\ref{fig:date} is the fraction of site lines that are matched to the rise or set
of a sacred celestial body vs date, and the combined p-value by date.

Under the assumption that the Sun Temple was used as an observatory of the celestial bodies considered
sacred to the Pueblo,
construction dates after 1250 AD appear to be more likely. However it must be noted that construction dates
before that time cannot be statistically excluded.

\section{Summary}

In our analysis, we have examined the potential that the
Sun Temple at Mesa Verde was used for astronomical
observation.
Previous studies have shown that the Sun Temple, as viewed
from structures at Cliff Palace, is aligned on the horizon to the set of the
Sun at winter solstice and the Moon at its major
southern standstill~\cite{malville1993prehistoric,munson2015mesa,munson2010reading}.  
However, ours is the
first analysis to examine potential astronomical alignments as viewed from within
the immediate vicinity of the Sun Temple site itself.

We have striven to ground our analysis in the known ethnography of the Pueblo peoples.
We observe a large, and statistically significant, number of site alignments to celestial
bodies known to be considered sacred to the Pueblo Indians, but do not observe a statistically significant
number of alignments to bodies {\it not} considered sacred.

In addition, examination of the number of matches to site lines to celestial bodies (and vice versa)
versus hypothesized year of construction of the temple reveals that the most likely construction date 
was 1250 to 1275 AD, precisely the time frame previously posited based on dendrochronological dates
from nearby structures like Cliff Palace that are assumed to be contemporaneous.

The placement of the kivas appears to have been carefully designed such that the two easternmost kivas were primarily involved 
in the observation of the setting of sacred celestial bodies, and the two westernmost kivas primarily were involved in
 observation of the
rise.

Two of the site alignments go through the feature of the Sun Temple known as the ``Sun Shrine'', and
thus the feature perhaps carried special significance to the builders of the complex.  These two 
alignments are within one degree of the declination of the rise of Pleiades and rise of Vega in 1250 AD, respectively.
It is interesting to note that the star-shaped eroded basin in
the Sun Shrine, with its eroded cupules in the center,
 may have been seen to represent the approximate appearance
of Pleiades, and thus the ``Sun Shrine'' may instead have actually been a shrine related to the Pleiades.

In our work we only considered alignments involving the four kiva-like structures in order to 
reduce, as much as possible, the number of site alignments considered, and also because
studies have shown that they appear to have been the first features constructed at the site~\cite{munson2010reading},
and thus likely played a key role in the ceremonial function of the structure.
However, the form of the Sun Temple is obviously much more complex than just the four kivas, 
and it is quite possible that other features of the Sun Temple were
aligned such that they too could be used for astronomical observations.
For instance, the lack of an alignment among the four kivas to the northern and southern rise or set of the
Moon at the major standstills does not mean that such an alignment does not exist involving
other architectural features of the Sun Temple;
to wit, the line going through the southwest
corner of the temple and tangent to the top of Kiva C is aligned with the rise of the Moon at the major northern standstill,
and the line going through the southeast corner and the center of Kiva C is
aligned with the rise of the Moon at the major southern standstill.
Further study is required to document all such potential astronomical alignments, and
assess their statistical significance.

\begin{figure}[h]
   \begin{center}
    \mbox{\put(-190,0){ \epsfxsize=13cm
           \epsffile{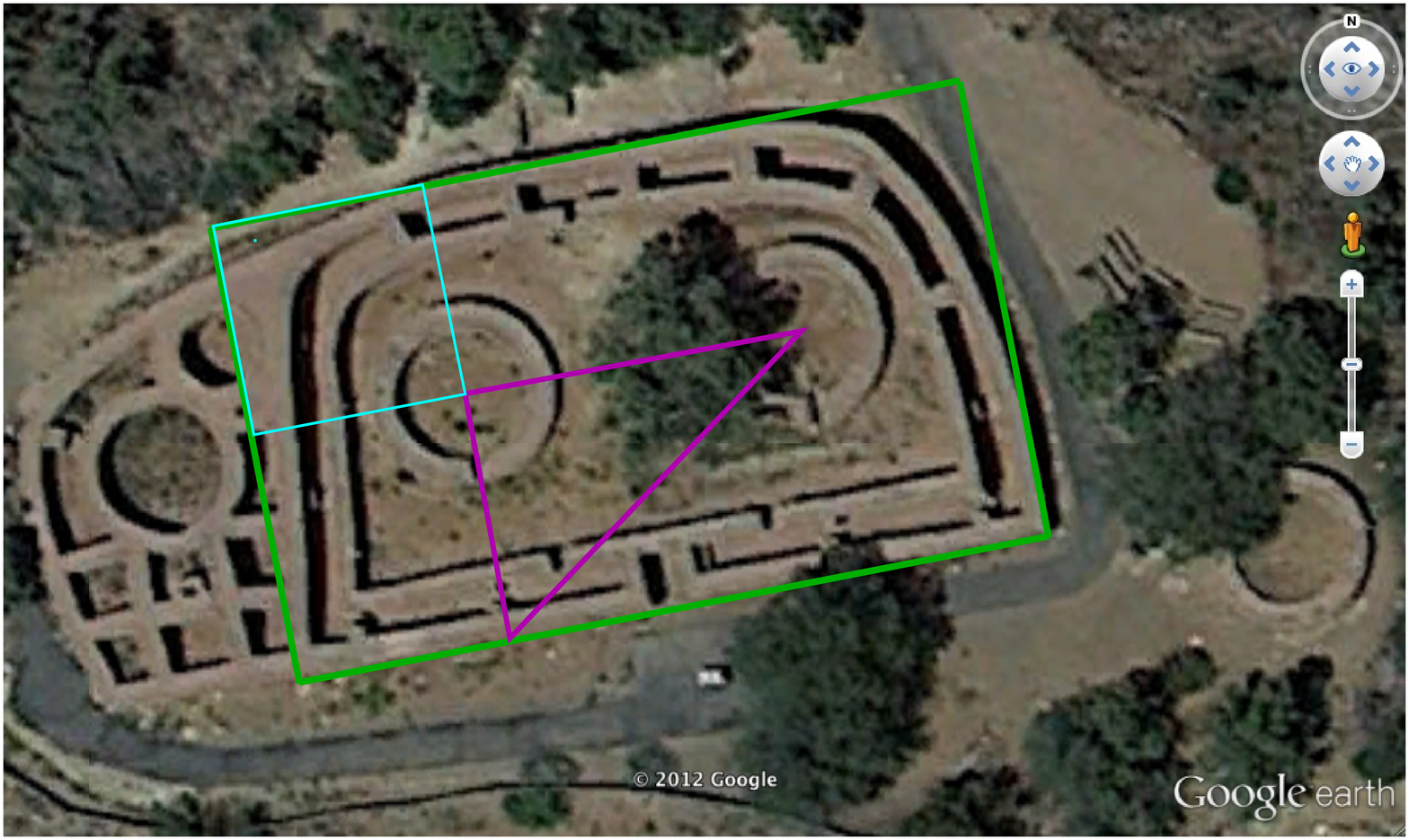}
     }}
     \vspace*{-0.0cm}
  \caption{
      \label{fig:shape}
A rectangle overlaid on the D-shaped structure has a ratio of length to width that is
within 1\% of the Golden Ratio.  
There is also compelling evidence that the architect was
aware of Pythagorean 3:4:5 triangles (for example, 
as indicated in magenta), and there is evidence of several squares in the
geometry of the complex (for example, as indicated in cyan).
Aerial view obtained from Google Earth
(accessed February 1, 2015). 
   }
   \end{center}
 \end{figure}

Beyond the utility of the Sun Temple site as an astronomical observatory, it cannot
be denied that it is pleasing to look at; for instance, the rectangle enclosing the large D has a ratio
of length to width that is within 1\% of the Golden Ratio
(see Figure~\ref{fig:shape}).  It seems unlikely that this is merely coincidental, and points
to a rather sophisticated knowledge of geometrical constructs.
In addition, there is compelling evidence that the architect was aware of Pythagorean 3:4:5 triangles
(see Figure~\ref{fig:shape}).
Previous studies have also shown evidence of equilateral and isosceles triangles in the
architectural features of the complex~\cite{munson2010reading}.
These symmetries, combined with the functionality of
the site, are the work not just of clever invention, but of unmitigated genius; even with the aid of
sophisticated computers, it would be extraordinarily difficult to design 
a site to have all the functionality that the Sun Temple achieves with just four simple circles,
let alone include a variety of geometrical constructs in the architecture.
The fact that it was designed without even the aid of
a written language or numeral system is truly amazing.

Several previous studies have shown that the Pueblo peoples in the northern San Juan and Chaco region
were engaged in detailed sky-watching~\cite{munro2011ancestors,mckim2015great,sofaer1981lunar,malville1993prehistoric}. 
However, it should be pointed out that, until now, these
studies have almost exclusively centered on site alignments to Sun solstices and lunar standstills, a
focus that quite likely is indicative of the ethnographic preferences of our own society, and
not necessarily that of the 
Pueblo peoples.  Indeed, the preferential focus of our society
on the study of potential solar and lunar alignments in ancient sites is not just concentrated
on sites in the American southwest, but sites all over the world.
The analysis methods we have 
developed in this study are applicable to any archaeological site in the
world whose architectural features are visible from the air, and it will be interesting
in the near future if these methods are applied to analyze the potential use of other ancient
sites as astronomical observatories.

\section{Acknowledgments}

The author is grateful to Dr.Mark Varien of the Crow Canyon Archaeological Center, 
and Dr.Wes Bernardini of the University of Redlands for useful and informative discussions related
to this work.

\begin{table}[t]
\caption{
\label{tab:tab1}
Celestial bodies considered sacred to the Pueblo peoples;
``line match'' is the 
site alignment, as enumerated on Figure~\ref{fig:matches}.
Deneb, Mintaka, and the 
Moon at its major northern and southern standstills
are the only four of the 28 celestial bodies considered sacred to the Pueblo that did not have 
an apparent rise or set match to one of the site lines drawn between the four kiva-like structures.
}
\begin{center}
\begin{tabular}{|l|r|r|r|c|} \hline
Name & Magnitude & Declination & Azimuth & Line Match \\
\hline
Sun equinox rise & $ -27 $ & $ 0 $ & $ 90.4 $ & $ 6 $ \\
Sun winter rise & $ -27 $ & $ -23.5 $ & $ 120.2 $ & $ 15 $ \\
Sun summer rise & $ -27 $ & $ 23.5 $ & $ 61 $ & $ 3 $ \\
Moon minor south rise & $ -13 $ & $ -18.4 $ & $ 113.5 $ & $ 13 $ \\
Venus northern rise & $ -4 $ & $ 23.7 $ & $ 60.6 $ & $ 3 $ \\
Venus southern rise & $ -4 $ & $ -23.6 $ & $ 120.1 $ & $ 15 $ \\
Arcturus rise & $ -0.05 $ & $ 23.2 $ & $ 61.3 $ & $ 3 $ \\
Vega rise & $ 0.03 $ & $ 38.3 $ & $ 40.3 $ & $ 1 $ \\
Spica rise & $ 0.98 $ & $ -7.1 $ & $ 99.1 $ & $ 10 $ \\
Antares rise & $ 1.06 $ & $ -24.4 $ & $ 121.3 $ & $ 15 $ \\
Pleiades rise & $ 1.6 $ & $ 21.5 $ & $ 63.2 $ & $ 2 $ \\
Alnilam rise & $ 1.69 $ & $ -2.0 $ & $ 92.7 $ & $ 8 $ \\
Alnitak rise & $ 1.74 $ & $ -2.6 $ & $ 93.5 $ & $ 8 $ \\
Saiph rise & $ 2.07 $ & $ -10.2 $ & $ 102.9 $ & $ 12 $ \\
Sirrah rise & $ 2.07 $ & $ 25.0 $ & $ 59.2 $ & $ 5 $ \\
Sadr rise & $ 2.23 $ & $ 38.0 $ & $ 40.7 $ & $ 1 $ \\
Scheat rise & $ 2.44 $ & $ 24.1 $ & $ 60.1 $ & $ 3 $ \\
\hline
Sun equinox set & $ -27 $ & $ 0 $ & $ -91 $ & $ 6 $ \\
Sun winter set & $ -27 $ & $ -23.5 $ & $ -121.2 $ & $ 5 $ \\
Moon minor north set & $ -13 $ & $ 18.4 $ & $ -67.6 $ & $ 14 $ \\
Rigel set & $ 0.18 $ & $ -9.3 $ & $ -102.7 $ & $ 7 $ \\
Betelgeuse set & $ 0.45 $ & $ 6.9 $ & $ -82.2 $ & $ 9 $ \\
Antares set & $ 1.06 $ & $ -24.4 $ & $ -122.2 $ & $ 5 $ \\
Bellatrix set & $ 1.64 $ & $ 5.4 $ & $ -84.2 $ & $ 11 $ \\
Alnilam set & $ 1.69 $ & $ -2.0 $ & $ -93.3 $ & $ 4 $ \\
Alnitak set & $ 1.74 $ & $ -2.6 $ & $ -94.1 $ & $ 4 $ \\
Sirrah set & $ 2.07 $ & $ 25.0 $ & $ -59 $ & $ 15 $ \\
Markab set & $ 2.49 $ & $ 11.3 $ & $ -76.7 $ & $ 12 $ \\
Algenib set & $ 2.83 $ & $ 11.0 $ & $ -77 $ & $ 12 $ \\
Albereo set & $ 3.05 $ & $ 26.6 $ & $ -56.9 $ & $ 16 $ \\
\hline
Moon major north rise & $ -13 $ & $ 28.6  $ & $ 54.8 $ & not matched \\
Moon major south rise & $ -13 $ & $ -28.6 $ & $ 127.1 $ & not matched \\
Deneb rise   & $ 1.25 $ & $ 42.8 $ & $ 33.3 $ & not matched \\
Mintaka rise & $ 2.25 $ & $ -1.2 $ & $ 91.8 $ & not matched \\
\hline
Moon major north set & $ -13 $ & $ 28.6  $ & $ -54.5 $ & not matched \\
Moon major south set & $ -13 $ & $ -28.6 $ & $ -128.1 $ & not matched \\
Deneb set   & $ 1.25 $ & $ 42.8 $ & $ -33.6 $ & not matched \\
Mintaka set & $ 2.25 $ & $ -1.2 $ & $ -92.3 $ & not matched \\
\hline
\end{tabular}
\end{center}
\end{table}



\end{document}